\documentclass[prl,twocolumn,showpacs,amsmath,amssymb]{revtex4}

\usepackage{times}
\usepackage{graphicx}

\newcommand{\beq}{\begin{equation}}
\newcommand{\eeq}{\end{equation}}

\newcommand{\ba}{\begin{eqnarray}}
\newcommand{\ea}{\end{eqnarray}}

\def\L{\Lambda}

\def\e{\epsilon}
\def\t{\theta}
\def\b{\beta}

\def\ve{\varepsilon}
\def\vt{\vartheta}

\def\mnras{Mon. Not. R. Astron. Soc.}
\def\aap{Astron. Astroph.}

\def\gs{\mathrel{\lower0.6ex\hbox{$\buildrel {\textstyle >}\over{\scriptstyle \sim}$}}}
\def\ls{\mathrel{\lower0.6ex\hbox{$\buildrel {\textstyle <}\over{\scriptstyle \sim}$}}}

\begin{document}

\title{The role of $\L$ in the cosmological lens equation}

\author{Mauro Sereno}
\email{sereno@physik.unizh.ch}

\affiliation{
Institut f\"{u}r Theoretische Physik, Universit\"{a}t Z\"{u}rich, Winterthurerstrasse 190, CH-8057 Z\"{u}rich, Switzerland
}

\date{September 23, 2008}

\begin{abstract}
The cosmological constant $\L$ affects cosmological gravitational lensing. Effects  due to $\L$  can be studied in the framework of the Schwarzschild-de~Sitter spacetime. Two novel contributions, which can not be accounted for by a proper use of angular diameter distances, are derived. First, a term $\delta \hat{\alpha}_\L = 2m b\L/3$ has to be added to the bending angle, where $m$ is the lens mass and $b$ the impact parameter. Second, $\L$ brings about a difference in the redshifts of multiple images. Both effects are quite small for real astrophysical systems, $\delta \hat{\alpha}_\L \ls 0.1 \mu$arcsec and $\Delta z_\mathrm{s} \ls 10^{-7}$.
\end{abstract}


\maketitle

The cosmological constant $\L$ plays a central role in gravitational physics and observational cosmology, with a fine-tuned $\L  \sim 10^{-52}\mathrm{m}^{-2}$ favored by large scale structure observations as a possible choice for dark energy \cite{pe+ra03}. $\L$ should take part in all kinds of gravitational phenomena and investigations have been performed on planetary systems \cite{isl+al,wri+al}, gravitational equilibrium of structures and discs orbiting rotating black holes \cite{bal+al06} . Actual upper bounds from stellar tests give $\L \ls 10^{-42}\mathrm{m}^{-2}$ \cite{wri+al}. 

The role of $\L$ in gravitational lensing is still debated. The cosmological lens equation is usually derived combining `local' results on light deflection in the very neighborhood of the lens, derived using asymptotically flat metrics, with considerations on light propagation in the nearly homogeneous regions among source, deflector and observer \cite{sch+al92}. Effects related to the background spacetime are based on the Friedmann-Robertson-Walker (FRW) spacetime in which the lens is embedded and can be seen as `global'. $\L$ and other cosmological fluids affect the measurement of angles at the observer \cite{ri+is07,bak+al07,sch08}. Such an effect is related to the background metric and can be embodied by the angular diameter distances \cite{ser08,par08,kh+po08}. It is still an open question if there are further local effects of $\L$. 
The fact that the differential equation for a light path in the Schwarzschild-de~Sitter (SdS) spacetime, i.e. the spherically symmetric Schwarzschild vacuum solution with a cosmological constant, can be written in a form that does not involve $\L$ \cite{isl+al}, differently from other dark energy models \cite{fin+al07}, suggests that any local effect should be small. I will show how a new deflection term, due to local coupling between $\L$ and the lens, shows up so that the corrected cosmological lens equation, relating the position angle of the images, $\vt$, and the angle $\b$ at which the source would be seen in absence of the lens, should be written as
\beq
\label{equ1}
\b   \simeq \vt - \frac{D_\mathrm{ds}}{D_\mathrm{s}} ( \hat{\alpha} +\delta \hat{\alpha}_\L ), \  \ \delta  \hat{\alpha}_\L =  \frac{\vt_\mathrm{E}^2}{2}  \left( \frac{D_\mathrm{d}}{r_\L}\right)^2 \vt,
\eeq
where $\vt_\mathrm{E}$ is the angular Einstein ring, $m$ is the lens mass, $r_\L (\equiv \sqrt{3/\L})$ is the outer horizon in the de~Sitter metric, and $D_\mathrm{d}$, $D_\mathrm{ds}$ and $D_\mathrm{s}$ are the angular diameter distances between the observer and the lens, the deflector and the source and the observer and the source, respectively. We take $G=c=1$ throughout. Cosmological distances make up for global effects whereas the bending angle $\hat{\alpha}$ describes local interactions. The contribution of $\L$ to the local deflection, $\delta \hat{\alpha}_\L$, is then a new local effect.

Together with a correction to the bending, $\L$ brings about also a small difference in the redshift of the images
\beq
\label{equ2}
\Delta z_\mathrm{s}^\mathrm{Obs}  \simeq \frac{1+ z_\mathrm{s}}{1+ z_\mathrm{d}} \left( \frac{1}{ z_\mathrm{d}}- \frac{1}{ z_\mathrm{s}} \right)^{-1} \frac{\vt_+^2 -\vt_-^2}{2},
\eeq
where $z_\mathrm{d}$ and $z_\mathrm{s}$ are the redshift of the lens and of the source in absence of $\L$, respectively. $\vt_+$  and $\vt_-$ are the position angles of the two images ($\vt_+^2 -\vt_-^2 \simeq \beta \sqrt{\beta ^2+4 \vartheta _\mathrm{E}^2}$).

The above results can be properly derived in the framework of the SdS metric \cite{adl+al65},
\beq
\label{sds1}
ds^2=  f_\L (r) dt^2- f_\L (r)^{-1}dr^2 -r^2 \left( d \theta^2- \sin^2 \theta d\phi^2 \right),
\eeq
where $f_\L (r) \equiv  1- 2 m/r  -\L r^2/3 $ and $m$ is the black hole mass. The SdS metric is a special case of the Mc Vittie metric, which provides an exact description of a point-like lens embedded in a FRW spacetime. In the SdS spacetime, the cosmic expansion is driven only by $\L$. The strong advantage of working with the SdS coordinates is that lightlike geodesics are very well known. We can then work in a well defined framework which already accounts for cosmic expansion and curvature and avoids the problem of matching local and global effects.

Due to spherical symmetry, photon trajectories can be restricted to the equatorial plane, $\t= \pi/2$. Let us consider an observer in $\{r_\mathrm{o}, \phi_\mathrm{o} =0 \}$, where $\phi_\mathrm{o}$ has been fixed without loss of generality, and a light source in $\{r_\mathrm{s}, \phi_\mathrm{s} \}$. The orbital equation of a light ray can then be written in terms of the first integral of motion $b$ as
\beq
\label{geo1}
\phi_\mathrm{s} = \pm \int \frac{dr}{ r^2} \left[ \frac{1}{b^2 }+\frac{1}{r_\L^2} - \frac{1}{r^2}  + \frac{2 m}{r^3} \right]^{-1/2} ,
\eeq
where the sign of the integral changes at the inversion points in the $r$-motion.  We consider the weak deflection limit, where the source and the observer lie in remote regions very far from the lens and photons pass by the lens center at a minimum distance which is much larger than the gravitational radius, i.e. $m/b \equiv \e_\mathrm{m} \ll 1$. In a cosmological scenario, $r_\mathrm{o} \sim r_\mathrm{s}  \ls  r_\L$. Furthermore, for a typical lensing system $b/r_\mathrm{o} \sim b/r_\mathrm{s} \sim \epsilon_\mathrm{m}$ \cite{ke+pe05}. Quantities of interest can then be expanded according to the parameters $\e_\mathrm{m}$ and $\e_\L \equiv b/r_\L$ . The expansion technique is similar to \cite{ser08} with the main difference that \cite{ser08} considered a local system well inside the outer horizon ($r_\mathrm{o}, r_\mathrm{s}  \ll r_\L$) and decoupled from the global expansion. For the sake of brevity, results are grouped up to a given formal order in $\epsilon$, collecting terms coming from any combination of the two expansion parameters \citep{se+de06,ser08}. Even if in our calculations different terms are kept apart, for a typical galaxy cluster lens with mass $\sim 10^{14}M_\odot$ and $b \sim 0.1~\mathrm{Mpc}$, $\e_\mathrm{m} (\sim 5\times 10^{-5})$ and $\e_\L (\sim 2\times 10^{-5})$ are actually of the same magnitude. The integral in Eq.~(\ref{geo1}) can then be solved approximately \cite{ke+pe05,se+de06,ser08}. For $b>0$, we get for the azimuthal deflection
\ba
\phi_\mathrm{s} &  = & -\pi -\frac{4 m}{b} + b \left(\frac{1}{r_\mathrm{s}} + \frac{1}{r_\mathrm{o}}\right)  
-\frac{15 m^2 \pi }{4 b^2}  - \frac{128 m^3}{3 b^3}  \label{geo3}  \\
& + &  \frac{b^3}{6}  \left(\frac{1}{r_\mathrm{s}^3}+\frac{1}{r_\mathrm{o}^3}\right)  - \frac{2 m b}{r_{\Lambda }^2}   - \frac{b^3}{ 2r_{\Lambda }^2}  \left( \frac{1}{r_\mathrm{s}}+\frac{1}{r_\mathrm{o}}\right)
+  {\cal{O}}(\e^4).  \nonumber
\ea
The angle $\hat{\alpha}_0 =4 m/b$ is the well known main contribution to the bending whereas terms $\propto (m/b)^i$ represent higher order corrections to the Schwarzschild lens. Geometrical terms are combinations of the $(b/r_i)^i$-factors and are related to image ($\vt  \sim b/r_\mathrm{o}$) and source positions; the $(b/r_\L)^i$-factors account for the outer horizon in the associated FRW spacetime. Finally, the term $\delta \hat{\alpha}_\L  \equiv 2 b m/r_\L^2$ describes the local coupling between the lens and $\L$. As for $\hat{\alpha}_0$, neither the source or the observer position enter in $\delta \hat{\alpha}_\L $. The product $\L m b$ is the lowest dimensionless combination built with the quantities describing the local interaction of the photon with the lens, i.e. $b$ and $m$, and the cosmological constant. In what follows, we will make the case that such a local coupling should be considered in the lens equation. 

In a cosmological scenario, observer, lens and source are receding. Position angles should be considered in the locally flat frame of reference of the moving observer, where measurements are actually performed. The apparent angular position of the image, i.e. the angle $\vt$ between the tangent to the photon trajectory at the observer and the radial direction to the lens can be expressed in terms of the tetrad components of the four momentum $P$ of the photon at the observer, $\cos \vt = P^{[r]}/ P^{[t]}$ \cite{ser08,se+de08}. Neglecting deviations from the Hubble flow, the motion has to be radial ($v^r= dr/dt \neq 0, d\phi/dt=0$). In the the SdS metric
\beq
\sin \vt = \frac{  \sqrt{1-{v^{[r]}}^2(r_\mathrm{o})}}  {1 - v^{[r]}(r_\mathrm{o}) \sqrt{1 - (b/r_\mathrm{o})^2  f_\Lambda (r_\mathrm{o}) } }  \frac{b}{r_\mathrm{o}} \sqrt{f_\L  (r_\mathrm{o}) } ,   \label{geo4}
\eeq
with $v^{[r]} = (-g_{rr}/g_{tt})^{1/2}v^r$. The radial motion of a comoving observer in SdS coordinates can be derived by referring to the corresponding Mc Vittie form, where an observer in the Hubble flow has constant spatial coordinates. By means of standard coordinate transformations \citep{nol99}, we get
\beq
v^{[r]}(r_\mathrm{o})=(r_\mathrm{o}/r_\L) (1-2m/r_\mathrm{o})^{-1/2}  .  \label{geo5}
\eeq
The associated reference spacetime without the lens is crucial in writing the lens equation, since distances as well as source position $ \b$ are defined there \cite{boz08}. By tuning the lens mass to zero we get the de~Sitter metric, one of the few cases in which the RW metric can be put in a static form \citep{flo80}. We will consider the spatially flat RW model and the corresponding coordinate transformations. Distances can be easily computed in the associated RW spacetime, and then expressed in SdS coordinates. Since the azimuthal coordinate of the source is not known a-priori, we have to assume the source to be aligned with the line of sight from the observer to the lens. The angular diameter distance between a comoving source at $z_2$ and a comoving observer at $z_1$ is $D_{12}= r_\L (z_2-z_1)/(1+z_2)$. $D_\mathrm{d}$, $D_\mathrm{ds}$ and $D_\mathrm{s}$ can then be written in terms of radial coordinates plugging in the corresponding redshifts in the associated spacetime, $z_\mathrm{d} =r_\mathrm{o}/r_\L$ and $z_\mathrm{s} = (r_\mathrm{o}+r_\mathrm{s})/(r_\L -r_\mathrm{s})$. The angle $\b$ is also defined in the associated spacetime. In analogy with Eq.~(\ref{geo4}), $\b$ is written in terms of a fictitious constant of motion which solves the geodesic motion in Eq.~(\ref{geo1}) for the actual source and observer coordinates but for $m=0$ \cite{ser08}. The lens equation is then obtained by writing $\phi_\mathrm{s}$ as a function of either $\vt$ or $\b$ and equating the two expressions.

As far as angles are concerned, a natural expansion parameter can be based on the Einstein radius, $\varepsilon \equiv  \vt_\mathrm{E}/(4D)$, with
$\vt_\mathrm{E} \equiv \sqrt{ 4 m D /D_\mathrm{d}}$ and $D \equiv D_\mathrm{ds}/D_\mathrm{s}$ \cite{ke+pe05,se+de06}. Once we expand the lens equation as a series in $\ve$, the solutions take the form $\vt  \simeq  \vt_\mathrm{E} \left\{ \t_0 +\t_1 \ve + \t_2 \ve^2 \right\}$ \cite{ke+pe05,se+de06,ser08}. Up to including terms of order of ${\cal O} (\ve^2)$, $\Lambda$ enters only through the cosmological distances and the image positions $\vt$ solve the standard lens equation,
\beq
\label{lens5}
\b   \simeq \vt - D \hat{\alpha}, \ \ \ \hat{\alpha} \simeq  \frac{4m}{b_0} +\frac{15\pi}{4}\frac{m^2}{b_0^2} ,
\eeq
where the bending angle is the Schwarzschild one up to ${\cal O} (\ve^2)$ and $b_0 (\equiv D_\mathrm{d} \vt)$ is the approximated impact parameter. 

Gravitational coupling effects between the central mass and $\L$ show up at the next order, giving rise to additional contributions to the deflection that can not be accounted for by using angular diameter distances. In order to illustrate the effect of $\L$ while still keeping expressions simple, let us consider a source aligned to the line of sight ($\b =0$). In this symmetric configuration, a critical tangential circle shows up in the observer's sky instead of two images , with an angular radius of
\ba
\vt_\mathrm{t} &  =  & \vt_\mathrm{E} \left\{  1+ \frac{15\pi}{32} \ve + \left[ 4 -\frac{4 D^2}{3} -\frac{675 \pi^2}{2048}  \right. \right. \nonumber \\
& + & \left. \left. \frac{1}{4} \left( 1-\frac{1}{D}\right)\frac{1}{r_{\L \ve}^2 } + \frac{1}{r_{\L \ve}} \right] \ve^2 \right\} \label{lens7},
\ea
where $r_{\L \ve} \equiv r_\L/(4 D D_\mathrm{d})$. At this order, $\Lambda$ affects the image position. The term $\delta \vt_\mathrm{t}^\L =\vt_\mathrm{E}  \ve^2/( 4r_{\L \ve}^2)= (1/4) (D_\mathrm{d}/r_\L)^2 \vt_\mathrm{E}^3$ comes directly from the azimuthal deflection.

Since measured image positions depend on the observer motion, one might as well consider an observer comoving in the associated RW spacetime, $v^{[r]}=r_\mathrm{o}/r_\L$, or even other radial peculiar motions. The critical circle corresponding to a generic radial velocity $v^{[r]} \simeq (r_\mathrm{o}/r_\L) (1+ \delta v^{(2)} \ve^2)$ forms at
\ba
\vt_\mathrm{t} & \simeq & \vt_\mathrm{t} (\L =0)  +\left\{  \frac{1}{4} \left( 1-\frac{1}{2D}\right)\frac{1}{r_{\L \ve}^2 }  \right.  \label{lens8} \\
& +&  \left. \frac{1}{4 r_{\L \ve} (1-2 D r_{\L \ve})} \left( 1+  \frac{1-4 Dr_{\L \ve} }{2 D}\delta v^{(2)}\right)  \right\}  \ve^2 ,\nonumber
\ea
which reduces to Eq.~(\ref{lens7}) for a Mc Vittie comoving observer, $\delta v^{(2)} = 4D-1/r_{\L \ve}$. For a particular choice of $\delta v^{(2)}$, the peculiar velocity can cancel the effect of $\L$. Whereas some contributions to the radius depend on the choice of the radial motion, $\delta \vt_\mathrm{t}^\L$ does not.

The choice of the angular diameter distance might hide some other effects. What an observer really measures is the redshift of the source $z_\mathrm{s}$, which is then plugged in the FRW expression for the distance. The very general formula for the redshift is $1+ z_\mathrm{s} = g_{\alpha \beta} k_\mathrm{s}^\alpha U_\mathrm{s}^{\beta}/ g_{\alpha \beta} k_\mathrm{o}^\alpha U_\mathrm{o}^{\beta}$, with $k_\mathrm{o}^{\alpha}$  and $k_\mathrm{s}^{\alpha}$ the wavevectors of the light ray at the observer and at the source, respectively, and $U_\mathrm{o}^{\alpha}$  and $U_\mathrm{s}^{\alpha}$ the four-velocities of the observer and of the source, respectively. Assuming Mc Vittie comoving players, we get
\beq
z_\mathrm{s}^\mathrm{Obs} = \frac{f_\L (r_\mathrm{o}) }{f_\L (r_\mathrm{s}) } \frac{  \sqrt{1-\frac{2 m}{r_\mathrm{s}}} + \frac{r_\mathrm{s}}{r_\L} \sqrt{1- \frac{b^2}{r_\mathrm{s}^2}f_\L (r_\mathrm{s}) }  }{  \sqrt{1-\frac{2 m}{r_\mathrm{o}}}-\frac{r_\mathrm{o}}{r_\L}  \sqrt{1-\frac{b^2}{r_\mathrm{o}^2}f_\L (r_\mathrm{o})} } -1 . \label{red1}
\eeq
The dependence on the impact factor, which would disappear in absence of $\L$, makes the redshifts of the two images different. The difference can be written by expressing $b$ as a function of of $\vt$ and then expanding. In terms of redshifts of the associated spacetime, $\Delta z_\mathrm{s}^\mathrm{Obs}$ takes the form of Eq.~(\ref{equ2}). Such redshift effect depends on the light ray directions at the source and at the observer and is not linked to the total travel time delay. For $\b \sim \vt_\mathrm{E}$, the redshift difference is proportional to the square of the Einstein radius ($\propto m$). The effect is really small. For $\b \sim \vt_\mathrm{E}$, $r_\mathrm{o} \sim r_\mathrm{s} \sim r_\L/2$ and a galaxy cluster lens with mass $\sim 10^{15}M_\odot$, $\Delta z_\mathrm{s} \sim 10^{-7}$. 

Up to now, we have written distances in terms of redshifts of the associated spacetime. You might ask if measured redshifts could play a role. The light source and the observer are both massless in our model. Due to gravitational redshift, the measured redshift of the deflector will also differ from the associated $z_\mathrm{d}$. However, redshift measurements are based on spectra which integrate on all the emitting regions of the lens along the line of sight. Since we do not  know the effective radial coordinate we should use in the evaluation of the redshift, it is then safe to still consider for the redshift the associated value. The angular diameter distances based on the measured redshifts, and the corresponding Einstein radius $\vt_\mathrm{E}^\mathrm{Obs} \equiv ( 4 m D(z_\mathrm{d}, z_\mathrm{s}^\mathrm{Obs} )/ [ D(0, z_\mathrm{d}) D(0,z_\mathrm{s}^\mathrm{Obs} ) ])^{1/2}$, differ from the expressions based on the associated ones but anyway $\vt_\mathrm{E}^\mathrm{Obs}$ does not embody the $\L$-correction due to the azimuthal deflection. 

As a further check we could consider static observers in the SdS spacetime, $dr/dt=0$. In this case the distances in terms of radial coordinates are the same used in \cite[equations~(15-17)]{ser08}. Up to ${\cal O} (\ve^2)$, the lens equation has still the form of Eq.~(\ref{lens5}). Up to the next order, the critical circle forms at
\ba
\vt_\mathrm{t}^\mathrm{St}&  =  & \vt_\mathrm{E}^\mathrm{St} \left\{1+ \frac{15\pi}{32} \ve_\mathrm{St} + \left[ 4 -\frac{4 {D_\mathrm{St}}^2}{3} -\frac{675 \pi^2}{2048}  \right. \right. \nonumber \\
& + & \left. \left. \frac{1}{4} \frac{1}{{r_{\L \ve}}_\mathrm{St}^2 } - \frac{(1+16 {D_\mathrm{St}}^2 {r_{\L \ve}}_\mathrm{St}^2)^{1/2}  }{ 16 {D_\mathrm{St}}^2 {r_{\L \ve}}_\mathrm{St}^3   } \right] \ve_\mathrm{St}^2 \right\}  \label{lens13},
\ea
where the index $\mathrm{St}$ reminds that the distances to be used are those for the static case. The term $\delta \vt_\mathrm{t}^\L$ is still there. Equation~(\ref{lens13}) agrees with \cite{ser08} via a proper consideration of the different expansion scheme.

We have considered either static or moving observers in the SdS spacetime. For the moving case, we have considered observers  either comoving with the Hubble flow or with peculiar velocities.

We have considered static or moving observers, either comoving with the Hubble flow or with peculiar velocities. The common outcome is that the local coupling of $\L$ with the lens mass gives rise to an azimuthal shift whose effect can not be embodied by angular diameter distances. This has been verified considering either distances in the associated RW metric or distances based on the observed redshifts. Comparison of the above results makes it clear that the $\delta \hat{\alpha}_\L$ bending has a local origin and can be distinguished from other contributions, which vary with different assumptions on the distances and the radial motion and are connected to the presence of an outer horizon in the SdS spacetime. In a general $\L$CDM model of universe with dark matter, only the $\delta \hat{\alpha}_\L$ contribution should be retained, whereas other effects of the cosmological constant are already embodied by the angular diameter distances. That is why we end up with a lens mapping in the form of Eq.~(\ref{equ1}). The perturbed image positions are then
\beq
\vt \simeq \vt_{0} \left\{ 1 + \frac{D_\mathrm{d}^2}{2r_\L^2}  \frac{\vt_{0}^2}{1+ \vt_{0}^2/\vt_\mathrm{E}^2} \right\},
\eeq
with $ \vt_{0} = (\b \pm \sqrt{\b^2+4\vt_\mathrm{E}^2})/2$ the 0-th order solutions. The consequent correction to the critical angular circle is $\delta \vt_\mathrm{t}^\L = (1/4) (D_\mathrm{d}/r_\L)^2 \vt_\mathrm{E}^3$. The effect on the observed angles is really small, $\sim \vt_\mathrm{E}^3$. For a source at $z_\mathrm{s}=1$ behind a lens with $M \sim 10^{15}M_\odot$ at $z_\mathrm{d}=0.3$ in a standard $\L$CDM model, $\delta \vt_\L \sim 0.1 \mu$arcsec. Note that the local coupling gives rise to an attractive gravitational effect which can not be associated with the repulsive force due to a positive $\L$, whose effect is incorporated in the cosmological distances.

Whereas the SdS metric provides a proper framework for the spacetime near the lens, it can not reproduce the shear and focusing due to other matter inhomogeneities. Apart from the very neighbourhood of the lens, such lensing effects are sizeable and should be accounted for by using properly modified expressions for the distances \citep{sch+al92,kan98}. Cosmological fluids such as dark matter should contribute corrections to the deflection angle similar and opposite to $\L$ and one might be tempted to generalize our results by replacing $\L/3$ with the square of the Hubble constant, $H^2$, for a generic $\L$CDM model. However, differently from dark energy which is supposed to be homogeneously distributed, dark matter is highly clumped around collapsed object, so that we do not expect a local back-reaction of the kind of what we found for $\L$. Noteworthily, $\L$ is supposed to dominate the energy budget of the universe in the very far future so that the SdS spacetime is going to provide a very realistic description of the universe. 

Let us now briefly review some previous analyses prompted by a recent paper \cite{ri+is07}. Such results seem to be correct, with some apparent disagreement in the physical interpretation being only due to unphysical gauge effects. Even if the geodesic equations in either the SdS metric or the Schwarzschild metric are formally the same \cite{isl+al}, this does not imply that lensing phenomena are independent of $\L$ \citep{gib+al08}. Coordinate angles differ from observed angles \cite{ri+is07,sch08} and the effect of $\L$ can then be viewed as an additional contribution $\propto -\L b r_\mathrm{o}$ to the bending if the lens equation is written in terms of radial coordinates instead of angular diameter distances. However, such a contribution is not true local bending since it can be incorporated by the distances, see Eq.~(\ref{lens5}), as already shown in \citep[section 4]{ser08}, where a static observer in a local system was considered, in \cite{par08}, who perturbatively integrated the null geodesics in the Mc Vittie metric, and in \cite{kh+po08}, who considered quantities in RW coordinates. However, in \cite{ser08,kh+po08} the $\L m b$ term in the geodesic equation discussed here was considered of higher order and then neglected. On the other hand, \cite[see equation~(30)]{par08} likely missed such a term since he considered the distance $D_\mathrm{ds}$ to be much smaller than the horizon $r_\L$, an hypothesis that is correct for local systems but breaks down in a cosmological context. 

Approaches adopting the Einstein-Strauss method with positive $\L$, where the matter in a spherical region collapses to form the lens and the resulting SdS vacuole is matched into a FRW background, were also followed.  Assuming that once the light transitions out of the vacuole all the $\L$-bending stops, \cite{ish+al08} derived a contribution $-\L b r_\mathrm{v}/3$ to the bending, in which $r_\mathrm{v}$, the radial SdS coordinate of the vacuole boundary, replaces the coordinate of the observer $r_\mathrm{o}$. This contribution is related to the distance from the vacuole boundary to the lens center and do not spring from coupling effects so that it should be incorporated in the total distance $D_\mathrm{d}$ from the observer to the lens. Finally, \citep{sch08b} integrated the light motion piecewise in a flat FRW solution and in the SdS metric and then pasted the geodesics together at the vacuole radius. However, the lens equation was not provided and the contribution of $\L$ to the deflection was not singled out. Furthermore, some higher-order terms were dropped out in the integration in the SdS metric. 

Common sense suggests that a cosmological constant, which can not give rise to a preferential direction, can not make local bending by itself \citep{sim+al08}. Back-reaction with the lens can however brings about a correction to the deflection near the lens. Coupling terms of the kind of $H^2 r^2 \Phi $, with $\Phi$ being the Newtonian gravitational potential, show up in a perturbed RW metric. The $2 m b \L/3$ contribution to the bending found in this paper exploiting the SdS spacetime accounts for such coupling effects and is, together with the difference in the redshift of the images, a novel, even if small, feature of lensing. Such signatures are peculiar to the cosmological constant and their detection would allow to distinguish $\L$ from other forms of dark energy. Whereas astrometry at the $\mu$arcsec level could be performed by future planned observational facilities, the measurement of $\Delta z_\mathrm{s}$ seems even more challenging.

M.S. is supported by the Swiss National Science Foundation and by the Tomalla Foundation. M.S. thanks the Queen's University Belfast for the warm hospitality during the final stages of this work.

\end{document}